\documentclass[aps,prl,twocolumn,showpacs,floatfix,superscriptaddress]{revtex4-1}
\usepackage{graphicx}
\usepackage{color}
\begin{document}
\flushbottom
\title{Strong enhancement of superconductivity at high pressures within the charge-density-wave states of 2H-TaS$_2$ and 2H-TaSe$_2$}

\author{D. C. Freitas}

\affiliation{Univ. Grenoble Alpes, Inst. NEEL, F-38000 Grenoble, France}
\affiliation{CNRS, Inst NEEL, F-38000 Grenoble, France}

\affiliation{Centro Brasileiro de Pesquisas F{\'i}sicas, Rua Dr. Xavier
Sigaud, 150, Urca, Rio de Janeiro - RJ, Brasil}

\author{P. Rodi{\`e}re}
\affiliation{Univ. Grenoble Alpes, Inst. NEEL, F-38000 Grenoble, France}
\affiliation{CNRS, Inst NEEL, F-38000 Grenoble, France}

\author{M. R. Osorio}
\affiliation{Laboratorio de Bajas Temperaturas, Departamento de F\'isica de la Materia Condensada, Instituto de Ciencia de Materiales Nicol{\'a}s Cabrera, Condensed Matter Physics Center (IFIMAC), Universidad Aut\'onoma de Madrid, E-28049 Madrid, Spain}

\author{E. Navarro-Moratalla} 
\affiliation{Instituto de Ciencia Molecular (ICMol), Universidad de Valencia, Catedr\'atico Jos\'e Beltr\'an 2, 46980 Paterna, Spain}

\author{N. M. Nemes}
\affiliation{GFMC, Departamento de F\'isica Aplicada III, Universidad
Complutense de Madrid, Campus Moncloa, E-28040 Madrid, Spain}

\author{V. G. Tissen}
\affiliation{Institute of Solid State Physics, Chernogolovka, 142432 Moscow Region, Russia}
\affiliation{Laboratorio de Bajas Temperaturas, Departamento de F\'isica de la Materia Condensada, Instituto de Ciencia de
Materiales Nicol\'as Cabrera, Facultad de Ciencias \\ Universidad Aut\'onoma de Madrid, E-28049 Madrid, Spain}

\author{L. Cario}
\affiliation{Institut des Mat{\'e}riaux Jean Rouxel (IMN), Universit{\'e} de Nantes, CNRS, 2 rue de la Houssini{\`e}re, BP 32229, 44322 Nantes Cedex 03, France}

\author{E. Coronado} 
\affiliation{Instituto de Ciencia Molecular (ICMol), Universidad de Valencia, Catedr\'atico Jos\'e Beltr\'an 2, 46980 Paterna, Spain}
\author{M. Garc{\'i}a-Hern{\'a}ndez}
\affiliation{Instituto de Ciencia de Materiales de Madrid-CSIC, Cantoblanco E-28049 Madrid, Spain}
\affiliation{Unidad Asociada de Bajas Temperaturas y Altos Campos Magn\'eticos, UAM, CSIC, Madrid, Spain}
\author{S. Vieira}
\affiliation{Laboratorio de Bajas Temperaturas, Departamento de F\'isica de la Materia Condensada, Instituto de Ciencia de Materiales Nicol{\'a}s Cabrera, Condensed Matter Physics Center (IFIMAC), Universidad Aut\'onoma de Madrid, E-28049 Madrid, Spain}
\affiliation{Unidad Asociada de Bajas Temperaturas y Altos Campos Magn\'eticos, UAM, CSIC, Madrid, Spain}
\author{M. \surname{N\'u\~nez-Regueiro}}
\email[Corresponding author: ]{nunez@neel.cnrs.fr}
\affiliation{Univ. Grenoble Alpes, Inst. NEEL, F-38000 Grenoble, France}
\affiliation{CNRS, Inst NEEL, F-38000 Grenoble, France}
\author{H. Suderow}
\affiliation{Laboratorio de Bajas Temperaturas, Departamento de F\'isica de la Materia Condensada, Instituto de Ciencia de Materiales Nicol{\'a}s Cabrera, Condensed Matter Physics Center (IFIMAC), Universidad Aut\'onoma de Madrid, E-28049 Madrid, Spain}
\affiliation{Unidad Asociada de Bajas Temperaturas y Altos Campos Magn\'eticos, UAM, CSIC, Madrid, Spain}

\begin{abstract}
We present measurements of the superconducting and charge density wave critical temperatures (T$_c$ and T$_{CDW}$) as a function of pressure in the transition metal dichalchogenides 2H-TaSe$_2$ and 2H-TaS$_2$. Resistance and susceptibility measurements show that T$_c$ increases from temperatures below 1 K up to $8.5$ K at $9.5$~GPa in 2H-TaS$_2$ and $8.2$ K at $23$~GPa in 2H-TaSe$_2$. We observe a kink in the pressure dependence of T$_{CDW}$ at about $4$~GPa that we attribute to the lock-in transition from incommensurate CDW to commensurate CDW. Above this pressure, the commensurate T$_{CDW}$ slowly decreases coexisting with superconductivity within our full pressure range. 
\end{abstract}
 \date{\today} \maketitle

\section{INTRODUCTION}

The near absence of reports on superconductivity in graphene and related single layer systems is notorious, given the large efforts devoted presently to these materials. Only recently have superconducting signatures in doped graphene sheets\cite{Ludbrook15,Chapman15} and Ising superconductivity in trigonal prismatic monolayer MoS$_2$\cite{Lu2015,Suderow2015} and NbSe$_2$ \cite{Xi2015} been observed. Few-layer devices have been also made of transition metal dichalcogenides (2H-TX$_2$, with T$=$Nb,Ta and X$=$S,Se), which crystallize in a hexagonal arrangement of transition metal atoms separated by the chalcogen. Single molecular layers, formed by hexagonal T-X groups, show opposing tendencies with decreasing thickness---a reduction of superconducting T$_c$ in 2H-NbSe$_2$ and an increase in 2H-TaS$_2$\cite{Bana13,Cao15,Navarro15,Galvis13,Galvis14}.

Layered materials are generally very sensitive to modifications of their lattice parameters\cite{Wilson75, Berthier76, Coronado10,Feng2012,Morosan06,Sipos2008,Kusmartseva2009,Suderow05d,Klemm12}. To understand their superconducting T$_c$, we need to address the interplay between superconductivity and the charge density wave (CDW) and how this interplay evolves when varying the lattice parameters.

It has been argued \cite{Wilson75,CastroNeto01} that superconductivity and CDW strongly interact in the 2H-TX$_2$. Initial discussions pointed out that there was a mutually exclusive interaction. Indeed, T$_c$ decreases and T$_{CDW}$ increases when the the lattice parameter {a/c} ratio decreases as we pass from 2H-NbS$_2$ (T$_c=6$ K and T$_{CDW}=$ 0 K ), 2H-NbSe$_2$ (T$_c=7.2$ K and T$_{CDW}=$ 30 K), 2H-TaS$_2$ (T$_c=1$ K and T$_{CDW}=$ 80 K) and 2H-TaSe$_2$ (T$_c=0.1$ K and T$_{CDW}=$ 120 K). On application of pressure in 2H-NbSe$_2$, the CDW disappears above 5~GPa  and T$_c$ increases slightly up to 8.5 K at 10~GPa\cite{Berthier76,Suderow05d}, pointing out an exclusive interaction too. Measurements of the superconducting density of states and of vortex core shapes in 2H-NbSe$_2$ show that the superconducting gap of 2H-NbSe$_2$ is strongly shaped by the CDW and it has been argued that the CDW decreases the gap value along certain directions in real space\cite{GuillamonPRL}. Angular resolved photoemission also shows interesting correlations between superconducting and CDW Fermi surface features. If these are cooperative or exclusive is, however, not clearly established when taking different photoemission measurements into account\cite{KissScience,RossnagelPRB}. Thus, most experiments point out that, particularly from data in 2H-NbSe$_2$, the interaction seems to be of competing nature. The mutual interaction between superconductivity and magnetism is debated in the cuprate compounds, where cooperative interactions have been discussed\cite{Keimer15}. Here we find that, unexpectedly, CDW and superconductivity coexist in a large part of the phase diagram when applying pressure to the Ta based 2H-TX$_2$, namely 2H-TaS$_2$ and 2H-TaSe$_2$.

In the compounds with largest interlayer separation and highest CDW transition temperatures, 2H-TaS$_2$ and 2H-TaSe$_2$, the CDW and superconducting phase diagrams have been studied up to 4~GPa. The superconducting T$_c$ increases in both compounds, up to about 2.5 K in 2H-TaS$_2$ and 0.4 K in 2H-TaSe$_2$\cite{Smith75}. In 2H-TaS$_2$, the resistivity vs temperature as a function of pressure shows that the CDW appearing below 80 K at ambient pressure decreases down to about 66 K at 3.5~GPa in 2H-TaS$_2$\cite{Delaplace,Scholz82,Wilson75}. In 2H-TaSe$_2$, the ambient pressure phase diagram consists of an incommensurate CDW (ICDW) appearing at 120K and a lock-in transition at 90K to a commensurate CDW (CCDW). There is a reentrant lock-in transition with pressure\cite{McWhan80}. The incommensurate CDW occupies the whole temperature range above 2~GPa, but, above 4~GPa, it locks to the lattice and becomes again commensurate \cite{Chu76,Bulaevskii76,McWhan80,Littlewood82,Rice81,Steinitz81}. Here we study the effect of pressure up to 25GPa on  superconductivity and charge density waves in 2H-TaS$_2$ and 2H-TaSe$_2$ . We find that the CDW does not disappear up to the highest pressures studied  and that T$_c$ increases considerably up to close to $9$ K in both compounds.

\section{EXPERIMENTAL}

To make a comparative study of both T$_{CDW}$ and T$_{c}$ under pressure we measure the magnetic susceptibility and the resistivity of small samples. The samples were grown using vapour transport and the politype purity of the 2H phase was checked by powder X-ray diffraction\cite{Delaplace}. To this end, samples of synthesized crystals were ground and loaded inside a capillary ready for powder X-Ray diffraction (performed in ambient conditions). Indexation of the reflections of the powder pattern by assuming a hexagonal symmetry allowed for the identification of a single phase with the following unit cell parameters: a = b = 3.3137(2) \AA, c = 12.076(1) \AA for 2H-TaS$_2$ and a = b = 3.43910(5) \AA, c = 12.7067(2) \AA for 2H-TaSe$_2$. The Le Bail refinement of the room temperature powder pattern is in agreement with that described for the 2H phases of TaS$_2$ and TaSe$_2$ crystals. To measure the susceptibility, we use a diamond anvil cell with a pressure transmitting medium of a methanol-ethanol mixture ($4$:$1$), which is considered to yield quasi-hydrostatic conditions up to the pressures of interest in our experiment\cite{Suderow05d,Tateiwa10}. Pressure was determined by the ruby fluorescence method \cite{Horn89}. We measure on small single crystalline samples cut into parallelepipeds of size about $100\times100\times30\;\mu$m$^3$. The susceptibility is obtained by a conventional AC method using a transformer and a lock-in amplifier\cite{Suderow05d}. For the resistance we use a Bridgman pressure cell with steatite as the pressure transmitting medium   \cite{Garbarino07}. Platinum wires were passed through the pyrophillite gasket. Samples are cut into pieces of approximate size of $\sim$100$\times$400$\times$60$\times \mu$m$^3$ and contacted to the  platinum leads in the pressure cell. The temperature was controlled by a motor introducing gradually into the cryostat the cell attached to a cane.The electrical resistance measurements were performed using a Keithley 220 source and a Keithley 2182 nanovoltmeter. Two samples were measured simultaneously giving the same results. We could not appropriately determine the volume of samples nor the geometrical factor. Thus, we provide relative temperature variations of susceptibility and resistance.

Fig.\ 1 displays the susceptibility and resistance versus temperature curves obtained at different pressures at low temperatures. We determine T$_c$ from the onset of the superconducting resistive and magnetic transition curves, defined as the intersection of two tangents, one to the flat portion of the curve above and the second to the steepest variation in the signal below the superconducting transition. In all cases we obtain sharp transitions, providing an unambiguous determination of the superconducting T$_c$. Sometimes, we observe in the resistance measurements a small non-zero residual value in the superconducting phase, which we attribute to two contacts touching each other at one side of the sample in the pressure cell. This does not influence the determination of T$_c$.

\section{RESULTS}

The evolution of T$_{CDW}$ under pressure was determined by calculating the temperature derivative of the resistance, $dR(T) /dT$, from the measured resistance vs temperature curves . In Fig.\ 2, we show $dR(T) /dT$ for different pressures. The development of the CDW produces a gap on the Fermi surface that causes a sudden increase in the resistance which induces a downward peak in $dR(T) /dT$ \cite{HornGuidotti}. Their position in the curves are signalled on Fig 2 by small arrows. The obtained pressure dependence of T$_c$ and T$_{CDW}$ is shown in Fig.\ 3. 

\begin{figure}[h]
	\centering
		\includegraphics[width=.5\textwidth, bb = 0 250 500 610]{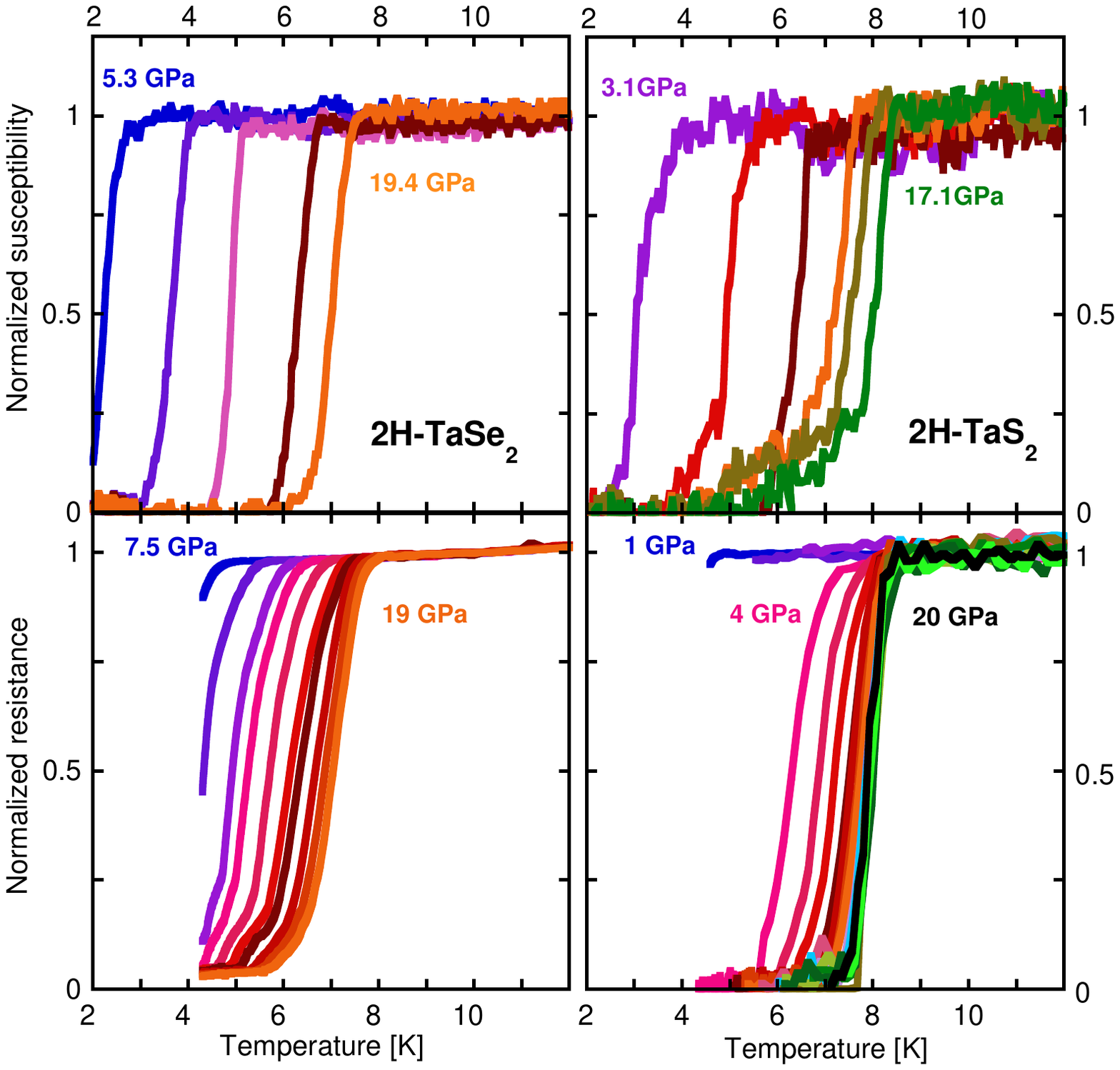}
	\caption{\footnotesize Upper panels:  Susceptibility of  2H-TaSe$_2$ (left panel) and 2H-TaS$_2$ (right panel) as a function of temperature for several applied pressures (5.3, 8.1 11.7 16.7 and 19.4~GPa for 2H-TaSe$_2$ and 3.1 5.8 7.9 9.8 14.8 and 17.1 GPa for 2H-TaS$_2$).	 Susceptibility has been normalized to the value found at 10 K, and the low temperature value has been modified to give zero. Lower panels: Resistance as a function of temperature for 2H-TaSe$_2$ (left panel) and for 2H-TaS$_2$ (right panel) for several applied pressures (from 7.5 to 15 by 1.5~GPa step and then every GPa up to 19 GPa for 2H-TaSe$_2$ and every GPa from 1 to 20 GPa for 2H-TaS$_2$. The resistance has been normalized to its value at 10 K.}
\end{figure}

\begin{figure}
	\centering
		\includegraphics[width=0.48\textwidth]{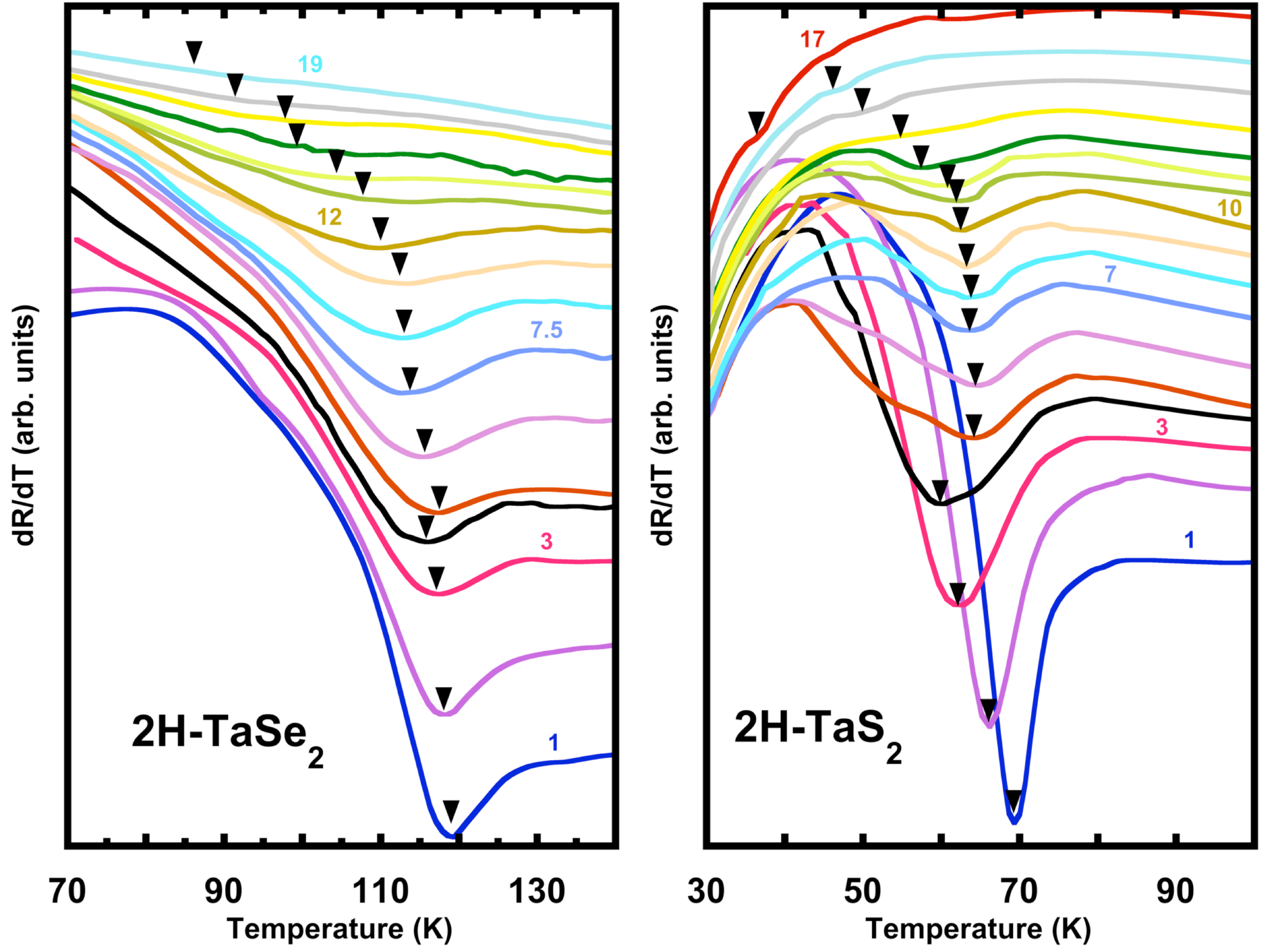}
	\caption{\footnotesize Derivative of the resistance close to the CDW ordering temperature for respectively 2H-TaSe$_2$ (left panel) and 2H-TaS$_2$ (right panel) for different pressures (1,2,3,4,5,6,7.5,9,10.5,12,13.5,15,16,17,18 and 19GPa (left) and every GPa from 1 to 17 GPa (right)). Curves are shifted in the y-axis for clarity. An arrow is used to mark the position where we take T$_{CDW}$ to give the pressure dependence discussed in Fig\ 3.}
\end{figure}

\begin{figure}
	\centering
		\includegraphics[width=0.48
	\textwidth]{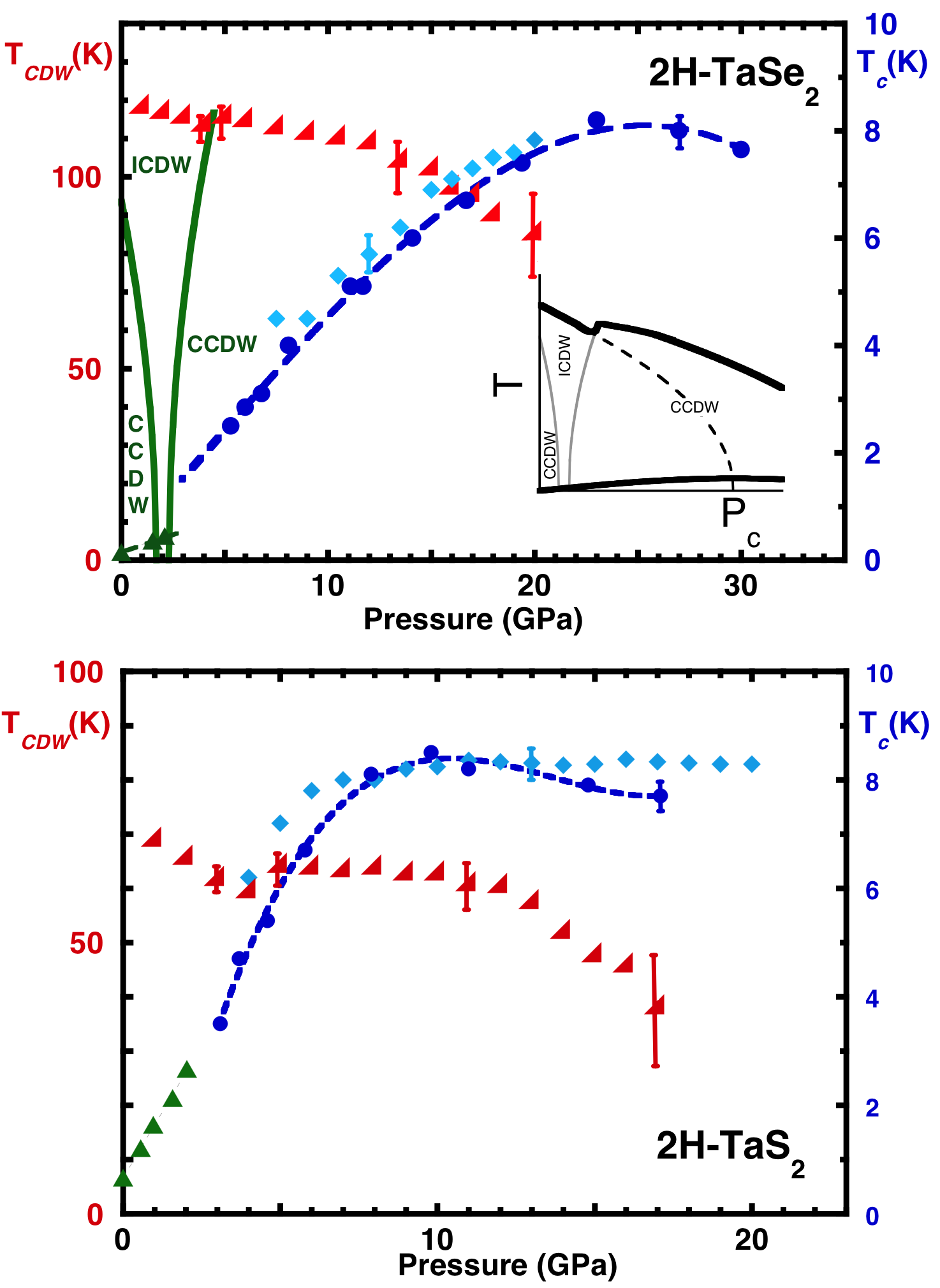}
	\caption{\footnotesize Phase diagram of the superconducting transition obtained from susceptibility (blue full circles), resistance (light blue losanges), and of the CDW transition (red triangles). Full triangles at low pressures have been obtained from Smith et al. \cite{Smith75}. Typical error bars are shown. Blue lines are guides to the eyes. Green dashed lines in the top panel are the positions of the CDW as found in previous works for 2H-TaSe$_2$\cite{McWhan80,Chu76,Delaplace}.  Insert: proposed phase diagram for both compounds. Black dashed lines are extrapolations of the low pressure range data (see text).}
\end{figure}

In 2H-TaSe$_2$, we find that pressure provokes an increase of T$_c$, with a slope of $0.58$ K/GPa, between $2$ and $8$ GPa. The maximum is attained at $T_c=8.2$ K at the pressure of 23~GPa. On the other hand, T$_{CDW}$, which signals, as discussed above, an ICDW transition, decreases slowly to about 4~GPa. At 4~GPa we observe a jump. On further compression, T$_{CDW}$ continues its rather slow decrease reaching a value of $ \sim{90}$K at 20GPa.

2H-TaS$_2$ shows a similar behavior. We find an initial slope $dT_c/dP=1.05$ K/GPa below 6~GPa, which then slows down to $0.45$ K/GPa, between 6 and 9.5~GPa. At low pressure, T$_{CDW}$ decreases slowly, similarly to previous results\cite{Delaplace} . At 4~GPa  we observe a sharp jump, similar to the one observed in 2H-TaSe$_2$. Above the jump, T$_{CDW}$ continues  its slow decrease down to $ \sim{40}$K at 17GPa. 

\section{DISCUSSION AND CONCLUSIONS}

Within the well-proven Bilbro-McMillan approach\cite{Bilbro}, the SC involves the portion of carriers which are not gapped by CDW, explaining their mutual competition. In quasi-1D systems, CDW's are originated by strong nesting of the parallel FS. Application of pressure destroys the CDW at a critical pressure P$_c$, where T$_c$ attains its maximum value\cite{Monceau12,Nunez92,Nunez93,Nunez13}.  While the loss of Fermi surface portions due to nesting dominates the interplay between CDW and superconductivity in quasi-1D systems, the situation is more involved in quasi-2D systems. Due to the quasi-cylindrical nature of the 2D FS's, the nesting anomalies are much weaker\cite{Rossnagel01} and then electron-phonon coupling is more important  in creating CDWs than in quasi-1D systems \cite{Bulaevskii76,Coleman88,Rossnagel11,Rice75,Johannes06,Johannes08}. The interplay between competing electronic and elastic degrees of freedom produces then minima in the free energy landscape that easily favor different kinds of CDW (commensurate or incommensurate) when modifying pressure and temperature.

 For 2H-TaSe$_2$, the jump in T$_{CDW}$ observed at about 4~GPa agrees with the previously reported pressure induced lock-in transition into a CCDW\cite{McWhan80}  (right green line in Fig.\ 3). Such a peak in the variation of density wave transition temperatures with pressure has been observed in other materials \cite{Klemme,Kaddour}  and have been unequivocally ascribed to transitions from an incommensurate to a commensurate state. McMillan explained the first order nature of the incommensurate / commensurate transition temperature driven by a Ginzburg Landau analysis \cite{McMillan77}. By analogy, we expect that the pressure driven incommensurate/commensurate transition  is also a first order transition and induce a jump in the T$_{CDW}$(P) phase diagram. 

The behaviour at ambient pressure for both compounds is similar. At high temperature they show a transition to an  ICDW, with a lock-in transition to an CCDW at lower temperatures \cite{McWhan80,Scholz82,Nishihara}. We can then speculate that the jump around 4GPa in both 2H-TaSe$_2$ and 2H-TaS$_2$ are due to the same phenomenon. That would imply that the ICDW in 2H-TaS$_2$ locks to the lattice with increasing pressure. We include this possibility in the proposed general phase diagram shown in the insert of Fig.\ 3.

The weak pressure dependence of T$_{CDW}$ at higher pressures indicates, on the other hand, that the CDW in this pressure range is remarkably robust to a reduction of the lattice parameters. This is not possible to explain within a pure nesting scenario, because band structure and the nesting condition are extremely sensitive to  pressure. \ {On other hand,  in the simplest lattice scenario through an e-ph coupling, theories have to reconcile the absence of pressure dependence of T$_{CDW}$ with the phonons hardening due to pressure.

It is interesting to discuss a recent model proposed to explain the CDW in the 2H transition metal dichalcogenides\cite{Gorkov12}. It considers that the CDW transition takes the form of a phase transition in a system of interacting Ising pseudo-spins. These can be  associated to the six transition metal atoms lying on the vertices of the in-plane hexagon described in Ref. \cite{Rossnagel11}. These might have a tendency to cluster in such a way as to form an inverse-star towards the transition metal atom at the center of the hexagon. This type of distortion has locally an intrinsic degeneracy, i.e. the transition metal atoms can choose between two inverse-stars with the same type of displacement.  By analogy with an up and down Ising ferromagnet, this is called a Ising pseudo-spin model. Thus, order is only short ranged and the development of a macroscopic static distortion corresponds to an ordering of the Ising pseudo spins\cite{nunez85}. This order-disorder transition is characterized by the absence of unstable zero energy phonon softening at the transition, as is confirmed by the reported non-zero soft mode for 2H-TaSe$_2$\cite{Moncton77} . Furthermore, the idea of pre-existing disordered deformations stems from the observation of a gap 5 times larger than the expected from weak-coupling formula\cite{Baker} and could be used to explain the robustness of the CDW at the higher pressures. It would imply, though, that the model describes better the transition metal dichalcogenides with large interlayer separation, 2H-TaSe$_2$ and 2H-TaS$_2$, than 2H-NbSe$_2$. In the latter compound the CDW has a lower critical temperature (30 K) and disappears already at 5~GPa\cite{Feng2012,Berthier76}.

Regarding the superconductivity, our measurements show that T$_c$ is strongly enhanced within the commensurate high pressure CDW phase. The increase in T$_c$ might be a consequence of phonon hardening or of Fermi surface induced changes with pressure. But it is not straightforward to think of a scenario where such features act on superconductivity independently to the CDW.  A possibility is that both phenomena involve widely different parts of the Fermi surface associated to the absence or small interband correlations.

It is interesting to note that the extrapolation of the low pressure (below the jump)  behavior of the ICDW up to higher pressures using a mean field approach $T_{ICDW}=T_{ICDW}^{0}\sqrt{\frac{P_{c}-P}{P_{c}}}$ (where $T_{ICDW}^{0}$ is the ambient pressure ICDW transition) leads to values for $T_{ICDW}^0$ becoming zero at a pressure $P_c$ roughly when the T$_c(P)$ curve ceases to increase in both materials. In the past, a mean field power law has been used for the low pressure commensurate transition in 2H-TaSe$_2$ \cite{Chu76,Freitas}. We have tentantively highlighted this aspect in the inset of Fig.\ 3. Although the extrapolation is, of course, connected with very large errors, it invites the speculation that there might be a mutually exclusive relation between incommensurate CDW and superconductivity. Note, however, that the Bilbro-McMillan approach discussed above in relation with the competition between incommensurate CDW and superconductivity does no longer apply when the incommensurate CDW has passed a transition into a commensurate CDW.

It is worth to note that the phase diagrams of 2H Ta-based dichalcogenides are in sharp contrast from that of other transition metal dichalcogenides such as 1T-TiSe$_2$, where superconductivity is observed at the vicinity of the pressure range of  ICDW \cite{Abbamonte,Kusmartseva2009} or 2H-NbSe$_2$, where T$_c$ is only moderately affected by the pressure\cite{Berthier76,Suderow05d}. In this last compound, the insensitivity of the superconducting critical temperature to the CDW transition is due to the fact that high energy optical phonon modes have a strong contribution to the Eliashberg function, whereas the
low-energy longitudinal acoustic mode that drives the CDW transition barely contributes to superconductivity\cite{Leroux15}. 

We conclude that understanding the value of T$_c$ in layered materials requires studying modifications of bandstructure, phonon dispersion and electron phonon coupling. Here we have shown that T$_c$ can considerably increase within a CDW, by more than an order of magnitude.

We acknowledge discussions with P. Grigoriev, R. Wehrt, N. Lera and J.V. Alvarez. D.C.F. gratefully acknowledges support from the Brazilian agencies CAPES and Cnpq. This work was partially supported by the French National
Research Agency through the project Subrissyme ANR-12-JS04-0003-01 and by the Spanish MINECO (MAT2011-22785 and FIS2014-54498-R), by the Comunidad de Madrid through program Nanofrontmag and by the European Union (Graphene Flagship contract CNECT-ICT-604391 and COST MP1201 action). The susceptibility vs temperature data at high pressures, where the strong increase in T$_c$ was first noticed, were taken by late V.G. Tissen of the Institute of Solid State Physics, Chernogolovka, Russia, during his sabbatical stay in Madrid financed by MINECO. We also acknowledge technical support of UAM's workshops, SEGAINVEX.

\end{document}